\documentclass[aps,twocolumn,superscriptaddress,amsmath,amssymb,longbibliography]{revtex4-1}

\usepackage[pdftex]{graphicx}
\usepackage[pdftex,bookmarks=true,bookmarksopen,bookmarksnumbered,
                colorlinks,
                linkcolor=blue,
                citecolor=blue]{hyperref}
\usepackage{dcolumn}
\usepackage{bm}
\usepackage{braket,amsmath}

\begin{document}
\title{Strain-induced spin resonance shifts in silicon devices}
\author{J.J. Pla}
\affiliation{School of Electrical Engineering and Telecommunications, University of New South Wales, Anzac Parade, Sydney, NSW 2052, Australia}
\author{A. Bienfait}
\affiliation{Quantronics Group, SPEC, CEA, CNRS, Universit\'{e} Paris-Saclay, CEA-Saclay, 91191 Gif-sur-Yvette, France}
\author{G. Pica}
\affiliation{Center for Neuroscience and Cognitive Systems @UniTn, Istituto Italiano di Tecnologia, Corso Bettini 31, 38068 Rovereto, Italy}
\affiliation{SUPA, School of Physics and Astronomy, University of St Andrews, KY16 9SS, United Kingdom}
\author{J. Mansir}
\affiliation{London Centre for Nanotechnology, University College London, 17-19 Gordon Street, London, WC1H 0AH, United Kingdom}
\author{F.A. Mohiyaddin}
\altaffiliation{Present Address: Quantum Computing Institute, Oak Ridge National Laboratory, Oak Ridge, TN 37830, USA}
\affiliation{School of Electrical Engineering and Telecommunications, University of New South Wales, Anzac Parade, Sydney, NSW 2052, Australia}
\author{Z. Zeng}
\affiliation{Universit\'{e} Grenoble Alpes, CEA, INAC-MEM, 38000 Grenoble, France}
\author{Y.M. Niquet}
\affiliation{Universit\'{e} Grenoble Alpes, CEA, INAC-MEM, 38000 Grenoble, France}
\author{A. Morello}
\affiliation{School of Electrical Engineering and Telecommunications, University of New South Wales, Anzac Parade, Sydney, NSW 2052, Australia}
\author{T. Schenkel}
\affiliation{Accelerator Technology and Applied Physics Division, Lawrence Berkeley National Laboratory, Berkeley, California 94720, USA}
\author{J.J.L. Morton}
\affiliation{London Centre for Nanotechnology, University College London, 17-19 Gordon Street, London, WC1H 0AH, United Kingdom}
\author{P. Bertet}
\affiliation{Quantronics Group, SPEC, CEA, CNRS, Universit\'{e} Paris-Saclay, CEA-Saclay, 91191 Gif-sur-Yvette, France}
\date{\today}

\begin{abstract}
In spin-based quantum information processing devices, the presence of control and detection circuitry can change the local environment of a spin by introducing strain and electric fields, altering its resonant frequencies. These resonance shifts can be large compared to intrinsic spin line-widths and it is therefore important to study, understand and model such effects in order to better predict device performance. Here we investigate a sample of bismuth donor spins implanted in a silicon chip, on top of which a superconducting aluminium micro-resonator has been fabricated. The on-chip resonator provides two functions: first, it produces local strain in the silicon due to the larger thermal contraction of the aluminium, and second, it enables sensitive electron spin resonance spectroscopy of donors close to the surface that experience this strain. Through finite-element strain simulations we are able to reconstruct key features of our experiments, including the electron spin resonance spectra. Our results are consistent with a recently discovered mechanism for producing shifts of the hyperfine interaction for donors in silicon, which is linear with the hydrostatic component of an applied strain.
\end{abstract}
\newpage
\maketitle

\section{Introduction}
The spins of dopant atoms in silicon devices have been shown to have great promise for quantum information processing (QIP) \cite{KanN98, FueNN12, PlaN13, TraAPL13, MuhNN14, FreQST17}. This has, in part, been encouraged by the extraordinarily long spin coherence times demonstrated, surpassing 1 second for the electron spin \cite{TyrNM12} and 3 hours for the nuclear spin \cite{SaeS13} of the phosphorus ($^{31}$P) donor. Another group-V donor with considerable promise for QIP in silicon is bismuth ($^{209}$Bi). Its large nuclear spin $I = 9/2$ and hyperfine constant $A = 1475$~MHz (which describes the interaction between the electron $\mathbf{S}$ and nuclear $\mathbf{I}$ spins $A\mathbf{S}\cdot\mathbf{I}$) provides rich features such as decoherence-suppressing atomic-clock transitions \cite{MohPRB10, WolfNN13, YasPRB16}, where coherence times can exceed by two orders of magnitude those typically achieved using other donor species. The Si:Bi system also possesses a large zero-field splitting of 7.375~GHz, making it an attractive dopant for use in hybrid superconducting devices \cite{BienNN15, BienN16} such as quantum memories \cite{WesPRL09, KuboPRL11, JusPRL13, GrePRX14}.

In donor-based QIP devices, such as quantum bits and hybrid quantum memories, the donors are located within close proximity of control and detection circuitry on the surface of the silicon chip. Recent experiments on individual donor electron and nuclear spin qubits adjacent to nanoelectronic circuits \cite{AngNL07} have highlighted the importance of considering the effect of these structures on the local environment of the spin. For example, it was shown that the spin resonance frequencies of $^{31}$P donors in nanoelectronic devices can experience shifts from their bulk-like values up to four orders of magnitude greater than their intrinsic line-widths \cite{PlaN12, MuhNN14, LauSA15, TraAPL16}. These shifts have been attributed to strain and electric fields produced by surface metallic gates in the devices.

Strain is an inherent feature of metal-oxide-semiconductor (MOS) electronic devices, which often combine materials that have vastly different coefficients of thermal expansion (CTE) \cite{LoNN15, ThoAIPA15}. It is therefore crucial to understand and predict the effect of intrinsic device strains on donors, as this can aid the design of scalable donor-based QIP and hybrid superconducting device architectures, serving as a guide to the often expensive and time-consuming fabrication process. Here we study a sample of bismuth ($^{209}$Bi) donors implanted from 50-150~nm beneath a thin-film aluminium wire (Figs.~\ref{fig:device}a and b). We observe the Si:Bi spin resonance spectra in the device to be substantially altered from what is typically found in bulk experiments \cite{WolfNN13,WeiAPL12}. Through analyzing a range of mechanisms, we conclude that strain induced by differential thermal contraction of the silicon and the surface aluminium structure is the most likely explanation for the non-trivial spectra. A model is developed that is able to reproduce many facets of our measurements, demonstrating the ability to predict device behavior and illustrating the importance of considering strain in semiconductor micro and nanoelectronic quantum devices.

The article is organized as follows: in Section~\ref{Sec:ES} we present the device architecture, physical system and experimental setup utilized in our study. Section~\ref{Sec:CFS} examines the electron spin resonance spectra of bismuth donors beneath an aluminium wire, revealing non-bulk-like splittings of the resonance peaks. Mechanisms potentially producing the splittings are discussed in Section~\ref{Sec:PSM} and simulations of the spin resonance spectra are performed in Section~\ref{Sec:SS} for one of the mechanisms identified. We conclude by discussing the implications of the simulations and the broader significance of our results for QIP in Section~\ref{Sec:S}.

\section{Experimental Details}\label{Sec:ES}
\subsection{Device}
\begin{figure}
\includegraphics[width=86mm]{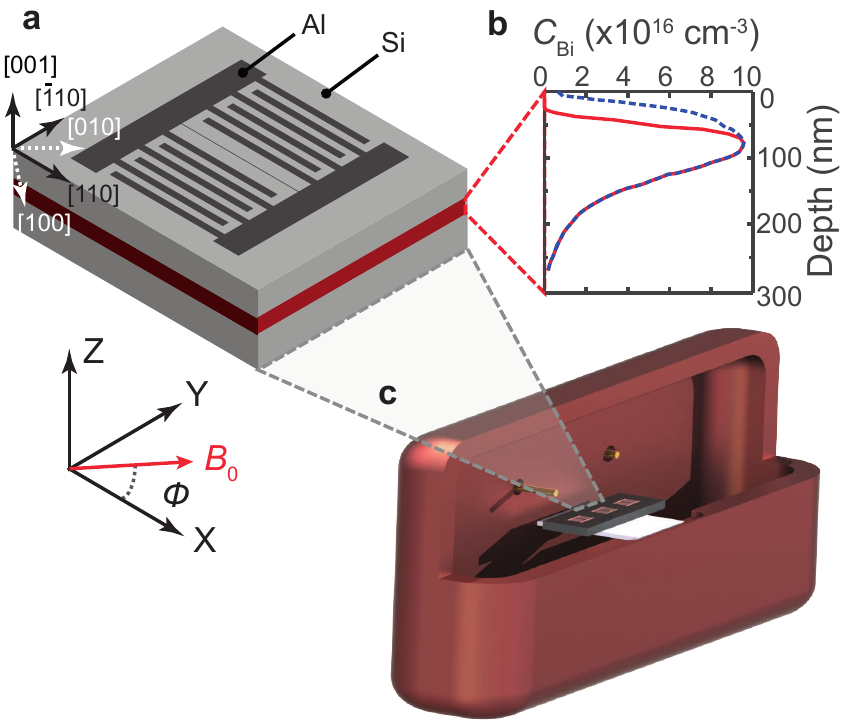}
\caption{\label{fig:device} \textbf{(a)} Sketch of an LC superconducting resonator made from a 50~nm thick film of aluminium, patterned on a silicon substrate, with central inductor 5~$\mu$m wide and 700~$\mu$m long. Whilst we only show one resonator here, there are three (almost identical) resonators patterned on the same chip (see panel c). The silicon sample was cleaved along the $\left<110\right>$ crystal axes and we specify a sample frame such that X~$\parallel \left[110\right]$, Y~$\parallel \left[\overline{1}10\right]$ and Z~$\parallel \left[001\right]$. The static field $B_{\rm 0}$ is oriented in the XY-plane at a variable angle $\phi$ to X. \textbf{(b)} Bismuth donor doping profile. The blue dashed curve shows the result of a secondary ion mass spectrometry (SIMS) measurement, whilst the red curve is the concentration of neutral donors obtained from a finite-element simulation performed using the SIMS profile, that takes into account donor ionization from the Schottky junction between aluminium and silicon (see Section~\ref{Sec:BIV}). \textbf{(c)} Three-dimensional copper microwave cavity sample holder. The silicon chip is mounted on a sapphire holder (pictured in white) and is probed via the cavity input and output antennas.}
\end{figure}

Our device (Fig.~\ref{fig:device}a) consists of three superconducting aluminium microwave resonators patterned on the surface of the same silicon chip via electron-beam-lithograph. The top 700~nm of silicon is an epitaxial layer of isotopically enriched 99.95\% $^{28}$Si, grown on a $\sim$~350 $\mu$m thick high-resistivity float-zone silicon (100) wafer. The epitaxial layer was implanted with $^{209}$Bi donors according to the profile depicted in Fig.~\ref{fig:device}b. 

The resonators are a lumped-element LC design, they contain a central inductive wire that produces an oscillating microwave magnetic field $B_{\rm 1}$ to drive and detect spin resonance. The drive field $B_{\rm 1}$ is proportional to the magnetic vacuum fluctuations $\delta B_{\rm 1}$ in the resonator, a quantity that we can simulate directly for our device. We utilize $\delta B_{\rm 1}$ in the following calculations and discussion: it is readily determined from our simulations (unlike $B_{\rm 1}$, which requires an accurate calibration of losses and other experimental parameters), and it provides us with another important measure, the spin-resonator coupling strength $g$. A simulation of $\delta B_{\rm 1}$ is performed knowing only the impedance of the resonator $Z_{\rm 0}$ and its frequency $\omega_{\rm 0}/2\pi$, and by calculating the resulting vacuum current fluctuations $\delta i = \omega_{\rm 0}\sqrt{\hbar/(2Z_{\rm 0})}$ in the wire (where $\hbar$ is the reduced Planck's constant). The current density distribution in the superconducting film (depicted in Fig.~\ref{fig:B1sim}a) is evaluated using DC equations adapted from Ref.~\cite{VanPTR99}, which are valid for the calculation of our microwave current due to the negligible ohmic losses at milli-Kelvin temperatures and because the typical resonator frequency ($\sim 7$~GHz) is significantly smaller than the superconducting gap of aluminium ($2\Delta(0) \approx 140$~GHz) \cite{BraPRB06}. The current density distribution is then fed to a finite-element magnetostatic solver (COMSOL Multiphysics), with the resulting $\vert\delta B_{\rm 1}\vert$ profile shown in Fig.~\ref{fig:B1sim}b. 

We observe a strong spatial dependence of the $\delta B_{\rm 1}$ orientation at the donor implantation depth (Fig.~\ref{fig:B1sim}c). Underneath the wire, the Y component of the field $\delta B_{\rm 1Y}$ dominates, whilst to the side $\delta B_{\rm 1Z}$ is the largest. We utilize this trait later in order to study spins in different spatial regions through orientation-dependent electron spin resonance (ESR) spectroscopy \cite{BienNN15, WisPRApp15}.

\subsection{Physical System}
At cryogenic temperatures, the bismuth donors bind an additional valence electron compared to the silicon atoms of the host crystal, providing a coupled electron ($S = 1/2$) and nuclear ($I = 9/2$) spin system that is described by the Hamiltonian:
\begin{equation}
\label{eq:FullHam}
H_{\rm 0}/h = \gamma_{\rm e}\mathbf{B_{\rm 0}}\cdot\mathbf{S} -  \gamma_{\rm n}\mathbf{B_{\rm 0}}\cdot\mathbf{I} + A\mathbf{S}\cdot\mathbf{I}
\end{equation}\\
where $\gamma_{\rm e} = 28$~GHz/T ($\gamma_{\rm n} = 6.963$~MHz/T) is the electron (nuclear) gyromagnetic ratio and $B_0$ is a static magnetic field applied in the plane of the aluminium resonators -- with a variable angle $\phi$ relative to the inductive wire (see Fig.~\ref{fig:device}a) -- that allows us to fine-tune the spin transition frequencies of the $^{209}$Bi donors. 

At values of the magnetic field where the electron Zeeman frequency $E_{\rm z}/h = \gamma_{\rm e}B_{\rm 0}\lesssim A$, the eigenstates become strongly mixed in the electron-nuclear spin basis and are best described by the total spin $\mathbf{F} = \mathbf{I} + \mathbf{S}$ and its projection onto $B_{\rm 0}$, $m_{\rm F}$ \cite{MohPRB10}. We chose the frequencies of the resonators to be close to the Si:Bi zero-field splitting of 7.375 GHz in order to minimize field-induced losses in the superconducting films, achieving $\omega_{\rm 0A}/2\pi = 7.305$~GHz for resonator A, $\omega_{\rm 0B}/2\pi = 7.246$~GHz for resonator B and $\omega_{\rm 0C}/2\pi = 7.143$~GHz for resonator C. We therefore operate in the regime where $F$ and $m_{\rm F}$ are good quantum numbers and we describe states in the $\vert F, m_{\rm F}\rangle$ basis. In the following analysis and discussion we focus on resonators A and B -- those with frequencies closer to the zero-field splitting which we were able to study the most extensively. Table~\ref{table:trans} presents important parameters that characterize the low-field ($B_{\rm 0} < 7$~mT) spin resonance transitions that are probed in our experiments.

\begin{table*}[t]
\begin{tabular}{ |p{0.6cm}|p{2.9cm}|p{1.9cm}|p{1.9cm}|p{1.9cm}|p{1.9cm}|p{1.9cm}|p{1.9cm}|p{1.9cm}|   }
\hline
 \multicolumn{9}{|c|}{Resonator A, $\omega_{\rm 0A}/2\pi = 7.305$~GHz} \\
 \hline
\multicolumn{2}{|l|}{Transition}  & $\Delta F\Delta m_{\rm F}$ & $B_{\rm 0}$ (mT) & $M$ & $(df/dB_{\rm 0})/\gamma_{\rm e}$ & $df/dA$ & $df/dg$ (MHz) & $df/dQ$\\
 \hline
1A &  $\vert 4, -4\rangle \leftrightarrow \vert 5, -5\rangle$ & -1 & 2.86    &	0.47   & -0.90   &	5.00 & -36.0 & 2.45\\
2A &  $\vert 4, -4\rangle \leftrightarrow \vert 5, -4\rangle$ & 0 & 3.22   &	0.30   &	-0.80    &	5.00 & -35.9 & -19.1\\
3A &  $\vert 4, -4\rangle \leftrightarrow \vert 5, -3\rangle$ & 1 & 3.69    &	0.07  &	-0.69   &	5.00 & -35.8 & -35.9\\
4A &  $\vert 4, -3\rangle \leftrightarrow \vert 5, -4\rangle$ & -1 & 3.69    &	0.42	&	-0.69    &	5.00 & -35.8 & 6.14\\
5A &  $\vert 4, -3\rangle \leftrightarrow \vert 5, -3\rangle$ & 0 & 4.32    &	0.40	&	-0.59    &	5.00 & -35.7 & -10.6\\
6A &  $\vert 4, -3\rangle \leftrightarrow \vert 5, -2\rangle$ & 1 & 5.22  &	0.13   &	-0.49       &	5.00 & -35.4 & -22.6\\
7A &  $\vert 4, -2\rangle \leftrightarrow \vert 5, -3\rangle$ & -1 & 5.22   &	0.37	 &	-0.49   &	5.00 & -35.5 & 7.42\\
8A &  $\vert 4, -2\rangle \leftrightarrow \vert 5, -2\rangle$ & 0 & 6.60    &	0.46    &-0.38    &	5.00 & -35.0 & -4.54\\
 \hline
 \multicolumn{9}{|c|}{Resonator B, $\omega_{\rm 0B}/2\pi = 7.246$~GHz} \\
 \hline
\multicolumn{2}{|l|}{Transition}  & $\Delta F\Delta m_{\rm F}$ & $B_{\rm 0}$ (mT) & $M$ & $(df/dB_{\rm 0})/\gamma_{\rm e}$ & $df/dA$ & $df/dg$ (MHz) & $df/dQ$
\\
\hline
1B &  $\vert 4, -4\rangle \leftrightarrow \vert 5, -5\rangle$ & -1 & 5.20    &	0.47 	&	 -0.90    & 5.00 & -65.4 &	2.49\\
2B &  $\vert 4, -4\rangle \leftrightarrow \vert 5, -4\rangle$ & 0 & 5.88  &	0.31	   &	-0.79     & 5.00  &	-65.1 & -19.0\\
3B &  $\vert 4, -4\rangle \leftrightarrow \vert 5, -3\rangle$ & 1 & 6.74   &	0.07   &	-0.69    &  5.00 & -64.8 &	-35.7\\
4B &  $\vert 4, -3\rangle \leftrightarrow \vert 5, -4\rangle$ & -1 & 6.75   &	0.42   &	-0.69    &  5.00 &	-64.9 & 6.27\\
 \hline
\end{tabular}
\caption{Numerical calculations of the spin transition parameters for the Si:Bi system at the LC resonator frequencies listed. Parameters include: resonance field ($B_{\rm 0}$), transition matrix element ($M = \vert\langle F, m_{\rm F}\vert S_{\rm X, Z}\vert F^\prime, m_{\rm F}^\prime\rangle\vert$ for $\vert\Delta F\Delta m_{\rm F}\vert = 1, 0$ transitions) and transition frequency sensitivity to: magnetic field ($df/dB_{\rm 0}$), electron $g$-factor ($df/dg$), hyperfine interaction ($df/dA$) and quadrupole interaction ($df/dQ$).}\label{table:trans}
\end{table*}

\subsection{Sample Mounting}
The device is fixed to a sapphire wafer with a small amount of vacuum grease (this serves to minimize sample strains produced through mounting) and the sapphire is then clamped between the halves of a rectangular copper microwave cavity (Fig.~\ref{fig:device}c), which acts as a sample enclosure and permits high quality-factors of the superconducting resonators by supressing radiation losses. The copper cavity is attached to the cold-finger of a dilution refrigerator and cooled to a base temperature of 20~mK, where we are able to detect the spin echo signals produced by the small number of shallow-implanted donors underneath each wire (estimated at $\sim 10^7$) by utilizing a quantum-noise-limited ESR setup, as described in Refs. \cite{BienNN15, BienPRX17, ProbAPL17}. We direct readers to the Supplementary Material of Ref.~\citenum{BienNN15} for a full schematic of the experimental setup.

\section{Spin Resonance Spectra}\label{Sec:CFS}
\subsection{Echo-Detected Field Sweep}
\begin{figure}
\includegraphics[width=86mm]{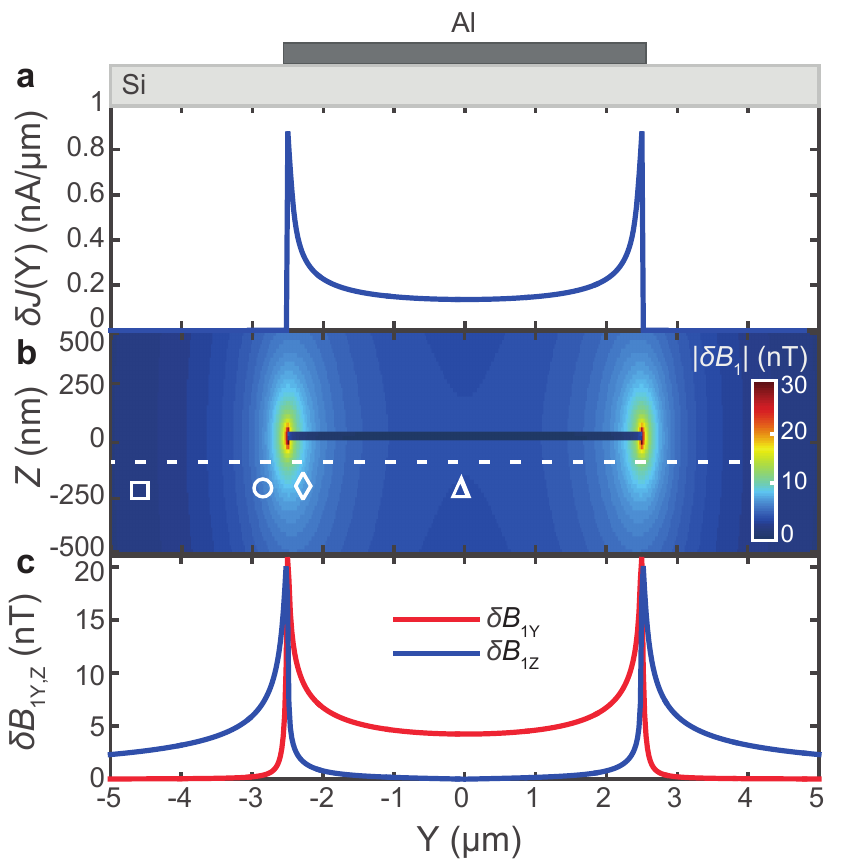}
\caption{\label{fig:B1sim} \textbf{(a)} Calculation of the current density vacuum fluctuations in the inductor. Equations describing the current density profile were adapted from Ref.~\cite{VanPTR99}. The only inputs to this calculation are the impedance of the resonator $Z_{\rm 0} = 44$~$\Omega$ and its frequency $\omega_{\rm 0}/2\pi \approx 7.3$~GHz, extracted using CST Microwave Studio. \textbf{(b)} A COMSOL Multiphysics finite-element simulation of the spatial dependence of the magnetic field vacuum fluctuations $\delta B_{\rm 1}$ magnitude produced by the current density in panel a. The symbols beneath the white dashed line identify regions that will be referred to in following sections. \textbf{(c)} Components of $\delta B_{\rm 1}$ along the Y and Z axes at a depth of 75~nm (corresponding to the peak donor concentration), marked by the white dashed line in panel b.}
\end{figure}

\begin{figure}
\includegraphics[width=86mm]{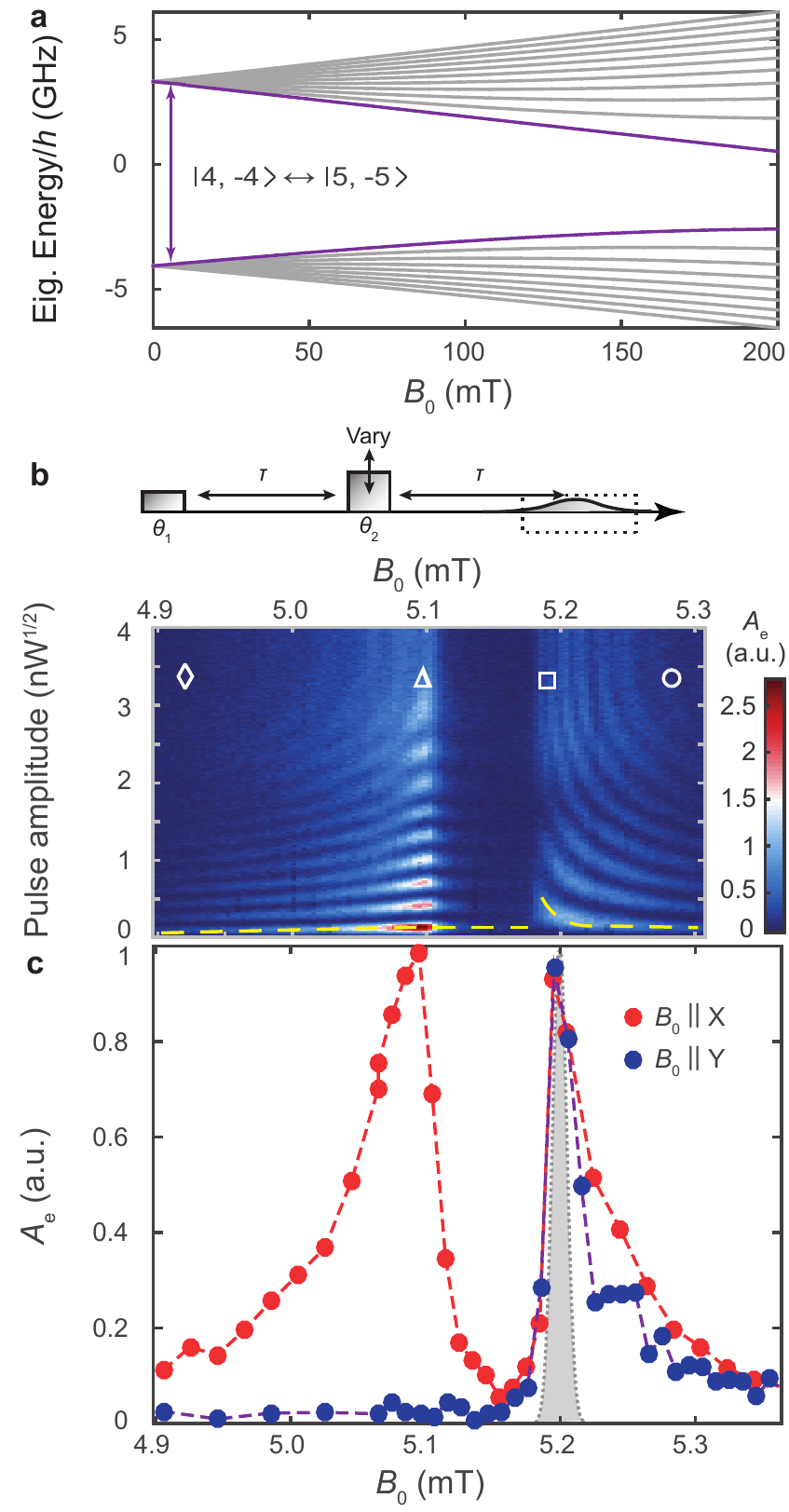}
\caption{\label{fig:compsweep} \textbf{(a)} Eigenstate frequencies of the Si:Bi system. The purple states and arrow identify the $\vert 4, -4\rangle \leftrightarrow \vert 5, -5\rangle$ transition (1B) probed in panels b and c. \textbf{(b)} Rabi oscillations as a function of $B_{\rm 0}$ for transition 1B. The amplitude of the refocusing pulse in a Hahn echo sequence (shown above) is varied to reveal oscillations in the integrated echo signal $A_{\rm e}$ (marked by the black dashed box in the sequence). Symbols identify spectral regions that are generated by spins at specific locations in the device (see Fig.~\ref{fig:B1sim}b). \textbf{(c)} A compensated echo-detected field sweep, taken using the calibrated $\pi$-pulse amplitudes of panel b (yellow dashed line). The grey-filled curve depicts the expected ESR spectrum, whilst the solid circles show the measured spectra (averaged over 8 sequences with a repetition rate of 0.2~Hz) for different field orientations. A 2\% correction was applied to $B_{\rm 0}$ for the measured data (within the magnet calibration error) so that the high-field peak aligns with the theoretical transition field. The same correction was applied to all experimental data in this study.}
\end{figure}

In this section we provide a detailed discussion of the Si:Bi ESR spectra, first reported in Refs.~\citenum{BienNN15, BienN16}. We observe the ESR spectrum for resonator B by performing an echo-detected magnetic field sweep on the lowest-field spin resonance line (indicated by the arrow in Fig.~\ref{fig:compsweep}a), corresponding to transition 1B, i.e. between the states $\vert 4, -4\rangle \leftrightarrow \vert 5, -5\rangle$ (see Table~\ref{table:trans}). We integrate the echo signal $A_{\rm e}$ from a Hahn echo sequence \cite{HahnPR50} (over the dashed region depicted in the pulse protocol of Fig.~\ref{fig:compsweep}b) and step the magnetic field $B_{\rm 0}$. The sweep is first performed with $B_{\rm 0} \parallel$~X ($\phi = 0^{\circ}$), and then repeated with the orthogonal orientation $B_{\rm 0} \parallel$~Y ($\phi = 90^{\circ}$); the resulting traces are shown in Fig.~\ref{fig:compsweep}c. The doped silicon sample investigated in this study has also been characterized using a standard ``bulk'' ESR spectrometer at X-band and with no planar on-chip resonator \cite{WeiAPL12}. The grey-solid curve in Fig.~\ref{fig:compsweep}c represents the spin resonance spectrum from this study extrapolated to the spin transition and frequency utilized in our experiment (see Appendix~\ref{app:LBM} for further details). Instead of measuring a single peak with a line-width of $\sim 20~\mu$T (as expected from the X-band measurement), we observe that the resonance is split into two peaks. Each peak has a line-width of $\sim 100~\mu$T, representing a total broadening of over an order of magnitude. 

Varying the amplitude of the refocusing $\pi$-pulse in the echo sequence reveals a series of Rabi oscillations ($A_{\rm e}$ is maximized whenever the refocusing pulse equals an odd-multiple of $\pi$) and the frequency of these oscillations is observed to depend strongly on the magnetic field $B_{\rm 0}$ (Fig.~\ref{fig:compsweep}b) across these two peaks. The traces in Fig.~\ref{fig:compsweep}c were recorded in a ``compensated'' manner, ensuring that at each value of $B_{\rm 0}$ the pulse amplitude was chosen to provide well-calibrated $\pi$ and $\pi/2$ pulses (yellow dashed line in Fig.~\ref{fig:compsweep}b).

The non-trivial peak splitting and field dependence of the Rabi frequency observed in Figs.~\ref{fig:compsweep}b and \ref{fig:compsweep}c can be understood by examining the experimental details, starting with the relevant bandwidths of the echo sequence. The $\pi/2$-pulse of the Hahn echo readout provides the initial excitation of spins that contribute to the echo signal. It has a duration of $t_{\rm \pi/2} = 2.5~\mu$s and an excitation bandwidth of $\sim~500$~kHz. This pulse is heavily filtered by the resonator, reducing its bandwidth to a value determined by the resonator line-width $\kappa = \omega_{\rm 0}/Q \approx 2\pi\times 25$~kHz (where $Q = 3.2\times 10^5$ is the quality factor of resonator B). Thus, only spins with resonant frequencies that lie inside the resonator bandwidth contribute to the measurement. In addition, these spins experience a relaxation rate which is three orders of magnitude greater than the intrinsic value, due to the Purcell effect \citep{BienN16}. This enhanced relaxation is suppressed quadratically with the spin-resonator frequency detuning, such that off-resonant spins display substantially longer energy relaxation times $T_{\rm 1}$ and quickly become saturated under the 0.2~Hz repetition rate of the experiment. Each $B_{\rm 0}$ in Figs.~\ref{fig:compsweep}b and \ref{fig:compsweep}c therefore corresponds to a highly-selective measurement on a small sub-ensemble of spins with a resolution $\Delta B = \kappa/(df/dB_{\rm 0})= 1~\mu$T, where $df/dB_{\rm 0}$ is the transition frequency field sensitivity (listed in Table~\ref{table:trans}). 

Comparing the echo-detected spectra for the different orientations of $B_{\rm 0}$ (red and blue circles in Fig.~\ref{fig:compsweep}c) provides strong evidence that the splitting and inhomogeneous broadening of the ESR transition results from the presence of the on-chip LC resonator. We find that the low-field peak vanishes for $B_{\rm 0} \parallel$~Y ($\phi = 90^{\circ}$) while the high-field peak remains relatively unchanged. This can be understood by referring to Fig.~\ref{fig:B1sim}c and noting that the spin transition probed here (1B, see Table~\ref{table:trans}) obeys the selection rule $\vert\Delta m_{\rm F}\vert = 1$ and is therefore excited only when $\delta B_{\rm 1} \perp B_{\rm 0}$. For $B_{\rm 0} \parallel$~Y, the condition $\delta B_{\rm 1} \perp B_{\rm 0}$ is only met for spins to the side of the wire (which experience a $\delta B_{\rm 1}$ field along Z). Spins underneath the wire (where $\delta B_{\rm 1}$ field almost entirely along Y) are not measured in this scan. For the spectrum recorded with $B_{\rm 0} \parallel$~X ($\phi = 0^{\circ}$), spins underneath the wire as well as those to the side observe $B_{\rm 0} \perp \delta B_{\rm 1}$ and thus contribute to the echo signal. Thus the low-field (vanishing) peak likely corresponds to the spins below the wire whilst the high-field peak is produced by spins to its side, indicating that the presence of the inductive wire is the source of the splitting. In Section~\ref{Sec:PSM} we discuss a number of potential mechanisms (e.g. electric field, Meissner-induced magnetic field inhomogeneity and strain) through which this could occur. The spin resonance frequency of the donors therefore depends on their location relative to the wire. By measuring only a small fraction of the large inhomogeneously broadened transition at each $B_{\rm 0}$ field ($1~\mu$T against $\sim~200~\mu$T) in Fig.~\ref{fig:compsweep}c, we are effectively probing sub-ensembles of donors residing in specific locations in the device.

We now return to the $B_{\rm 0}$ dependence of the Rabi oscillations (Fig.~\ref{fig:compsweep}b) and demonstrate that the picture described above is in good agreement with this data. The coupling strength between each spin and the resonator is given by $g = \gamma_{\rm e}M\vert\delta B_{\rm 1\perp}\vert$, where $M$ is the ESR transition matrix element (see Table~\ref{table:trans}) and $\vert\delta B_{\rm 1\perp}\vert$ is the magnitude of the $\delta B_{\rm 1}$ component felt by the spin that is perpendicular to $B_{\rm 0}$. The Rabi frequency $\Omega_{\rm R}$ then has a linear dependence on the $\delta B_{\rm 1}$ field through the relation $\Omega_{\rm R} = 2 g\sqrt{\overline{n}}$, where $\overline{n}$ is the mean intra-cavity photon number (proportional to the input microwave power). For the high-field peak in the ESR spectra (originating from spins located to the side of the wire), the sharp transition at the low-field edge likely corresponds to spins far from the wire that are bulk-like in their behavior. Being far from the wire, these spins also experience a reduced $\delta B_{\rm 1}$ (see Fig.~\ref{fig:B1sim}c) and thus Rabi frequency, observed as longer-period oscillations in Fig.~\ref{fig:compsweep}b. Moving closer to the wire increases the spin resonance shifts (i.e. through larger electric or strain fields) as well as the magnitude of the $\delta B_{\rm 1}$ field felt by the spins. We thus anticipate the tail regions of the lines to have an enhanced Rabi frequency, and this is indeed the case. The symbols overlaid on Fig.~\ref{fig:compsweep}b summarize the above discussion by correlating the different spectral regions with spins from specific locations in the device (see corresponding symbols in Fig.~\ref{fig:B1sim}b).



\subsection{Extended Spectra}
\begin{figure}[t]
\includegraphics[width=86mm]{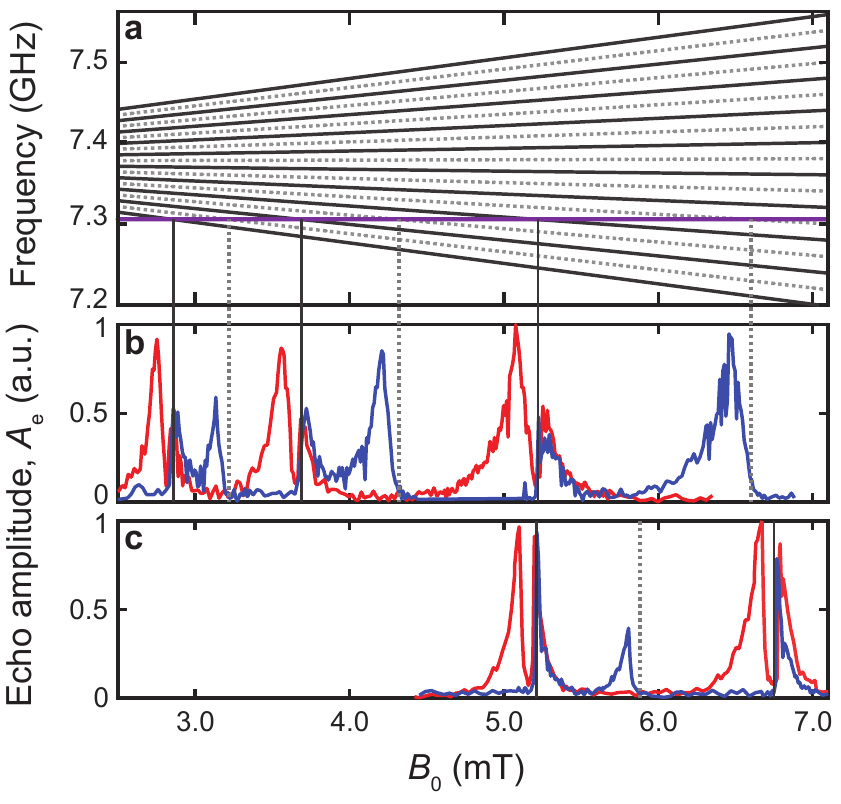}
\caption{\label{fig:extdspec} \textbf{(a)} ESR transition frequencies of the Si:Bi system for $B_{\rm 0} < 7$~mT. Solid lines represent the spin transitions that obey the selection rule $\Delta m_{\rm F} = \pm 1$ (i.e. $\delta B_{\rm 1}\perp B_{\rm 0}$) whilst the dashed lines show $\Delta m_{\rm F} = 0$ transitions ($\delta B_{\rm 1}\parallel B_{\rm 0}$). The purple solid line indicates the frequency of resonator A ($\omega_{\rm 0A}/2\pi = 7.305$~GHz). \textbf{(b)} Compensated echo-detected field sweeps of the ESR transitions below 7~mT of resonator A and \textbf{(c)} resonator B. The theoretical spin transition frequencies are identified by the black solid and dashed lines.}
\end{figure}

To help identify the mechanism behind the wire-induced peak splitting and broadening, we probe additional spin resonance transitions (listed in Table~\ref{table:trans}) using resonators A and B, which display different sensitivities to the various Hamiltonian parameters. In Fig.~\ref{fig:extdspec}a we plot the calculated low-field ESR transition frequencies and their crossing with resonator A. Transitions obeying the usual spin selection rule $\Delta m_{\rm F} = \pm 1$ (displayed in red) are accessed in the experiment by ensuring $B_{\rm 0} \perp \delta B_{\rm 1}$, as was the case for the previous measurement on transition 1B. Such transitions are typically referred to as being of ``$S_{\rm X}$'' type, since it is primarily the $S_{\rm X}$ operator that drives spin flips between the states. Transitions obeying the selection rule $\Delta m_{\rm F} = 0$ (blue lines in Fig.~\ref{fig:extdspec}a) -- of so-called ``$S_{\rm Z}$'' type -- are probed in the experiment with the alignment $B_{\rm 0} \parallel \delta B_{\rm 1}$ (i.e. $B_{\rm 0} \parallel Y$). We refer the reader to Appendix~\ref{app:SRT} for a detailed discussion of these two types of spin resonance transitions.

\begin{table}[b]
\begin{tabular}{ |p{2cm}|p{2cm}|p{2cm}| }
\hline
 \multicolumn{3}{|c|}{Resonator A, $\omega_{\rm 0A}/2\pi = 7.305$~GHz} \\
 \hline
Transition & $B_{\rm 0c}$ (mT) & $\Delta B_{\rm 0}$ (mT) \\
 \hline
1A & 2.87 & 0.11\\
4A &  3.71 & 0.15\\
7A &  5.28 & 0.20\\
 \hline
 \multicolumn{3}{|c|}{Resonator B, $\omega_{\rm 0B}/2\pi = 7.246$~GHz} \\
 \hline
Transition & $B_{\rm 0c}$ (mT) & $\Delta B_{\rm 0}$ (mT) \\
\hline
1B &  5.24 & 0.11\\
4B &  6.84 & 0.14\\
 \hline
\end{tabular}
\caption{Experimental center fields ($B_{\rm 0c}$) and peak splittings ($\Delta B_{\rm 0}$) extracted from the measured ESR transitions for resonator A and B. $\Delta F\Delta m_{\rm F} =  0$ transitions do not display a splitting and are therefore not included. Although the $\Delta F \Delta m_{\rm F} = \pm 1$ transitions are almost degenerate here, we attribute the peaks to the $\Delta F \Delta m_{\rm F} = -1$ transitions, which exhibit larger transition matrix elements $M$. We do not attempt to extract line-widths of the peaks due to their highly asymetrical shapes. }\label{table:extdspec}
\end{table}

A measurement of the first transitions with $B_{\rm 0} \parallel$~X ($\Delta m_{\rm F} = \pm 1$) is shown in the compensated echo-detected field sweep of Fig.~\ref{fig:extdspec}b (red trace) for resonator A. Also presented here is the spectrum recorded with $B_{\rm 0} \parallel$~Y (blue trace), which is composed of both $\Delta m_{\rm F} = 0$ resonances from spins underneath the wire (where $B_{\rm 0} \parallel \delta B_{\rm 1}$) and $\Delta m_{\rm F} = \pm 1$ resonances from spins to the side of the wire (where $B_{\rm 0} \perp \delta B_{\rm 1}$). The $\Delta m_{\rm F} = 0$ transitions are observed to lack a splitting, this is further evidence that they originate from spins located predominantly underneath the wire (the only region with $B_{\rm 0} \parallel$~Y). The experiments are repeated for resonator B and displayed in the lower traces of Fig.~\ref{fig:extdspec}c. We display the extracted peak splittings of the recorded transitions in Table~\ref{table:extdspec}.


\section{Peak Splitting Mechanisms}\label{Sec:PSM}
We now turn to the analysis of possible mechanisms through which the presence of the aluminium wire may induce a splitting and broadening of the observed ESR spectra.

\subsection{Built-in Voltage}\label{Sec:BIV}
\begin{figure}
\includegraphics[width=86mm]{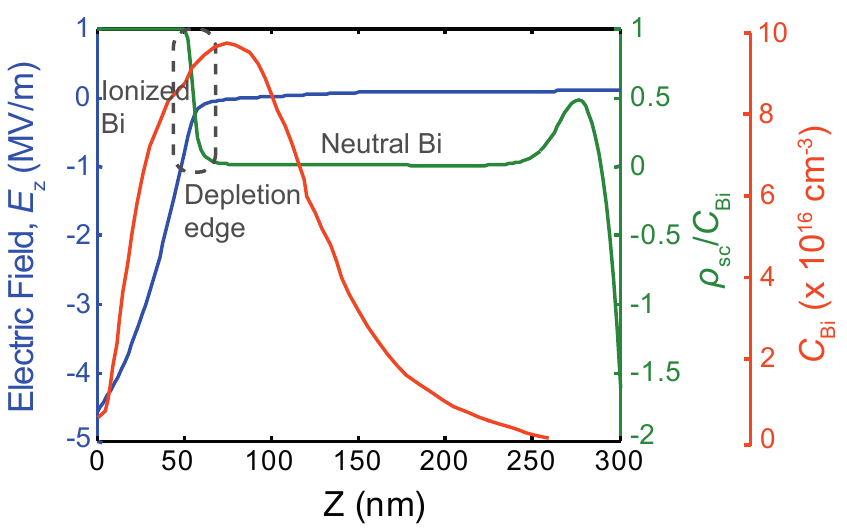}
\caption{\label{fig:builtin} Fraction of ionized donors, in-built electric field and doping profile versus depth beneath the aluminium resonator. The implanted Bi profile (as determined from a secondary ion mass spectrometry measurement) is shown in orange. The green curve provides the space charge density $\rho_{\rm sc}$ (which represents the density of ionized donors) divided by the donor concentration $C_{\rm Bi}$. Calculations were performed with the finite-element electrostatic solver ISE-TCAD, with a simulation temperature of 5~K (the minimum temperature at which convergence was achieved) and an assumed background boron doping density of $10^{13}$~cm$^{-3}$. The grey dashed box highlights the edge of the depletion region, this roll-off is expected to get steeper at the experimental temperature of 20~mK. At $\sim$~275~nm depth the bismuth donors are ionized once again (this time to the background boron acceptors present in the sample), before the space charge density becomes negative, indicating the presence of ionized boron dopants.}
\end{figure}

The aluminium/silicon interface formed beneath the resonator constitutes a Schottky junction. Band bending at the interface results from the difference in work functions of the aluminium and silicon (or from Fermi level pinning to surface states) \cite{RhoIEE82}. The band bending causes ionization of bismuth donors within an area known as the depletion region. Donor ionization continues into the semiconductor until a sufficient space-charge has been accumulated to counter the band bending. Immediately outside of the depletion region, the total electric field is reduced to zero. At finite temperatures, however, the edge of the depletion region is broadened according to Fermi-Dirac statistics, and a small fraction of neutral donors can experience large electric fields. Such donors would display a Stark shift of the hyperfine interaction \cite{LauSA15} or electron $g$-factor through the electric field, altering their resonant frequencies from those to the side of the wire away from the depletion region.

We have performed finite-element simulations with the commercial software ISE-TCAD, which solves the Poisson equation self-consistently to extract the electric fields and ionized bismuth concentration underneath the wire, the results of which are shown in Fig.~\ref{fig:builtin}. This plot demonstrates that the broadening of the depletion region edge is small relative to the width of the implantation profile, even at the elevated simulation temperature of 5~K --- the minimum temperature at which convergence was achieved. Donors at depths less than 50~nm are mostly ionized, whilst donors deeper than this are neutral and experience negligible electric fields ($< 50$~kV/m, with expected Stark shifts below 1~kHz \cite{PicPRB15}). At the experimental temperature of 20~mK, we expect an even sharper depletion region boundary. We therefore discount this mechanism as the cause for the spectral broadening and remove the shallow donors ($< 50$~nm) beneath the wire from the spectra simulations in the following sections.

\subsection{Magnetic Field Inhomogeneity}
It is conceivable that the superconducting resonator could perturb the static magnetic field in a manner that produces differing magnetic field profiles beneath the wire and to its side. For example, this might result from the component of a misaligned $B_{\rm 0}$ field perpendicular to the aluminium film, concentrating above or below the wire due to the Meissner effect \cite{UndAPS17}. The strength of any such inhomogeneity increases in proportion with the magnitude of $B_{\rm 0}$, and as the resonators are fabricated within 2~mm of one another on the same silicon chip, the inhomogeneity would be nearly identical for each of the resonators. We can rule this mechanism out due to the fact that we observe the same splitting and line-width of the first spin transition for resonators A and B (see Table \ref{table:extdspec}) and also C (see Appendix~\ref{app:MFI}), despite the transition for resonator C occurring at almost twice the field of resonator B and three times that of resonator A. 

\subsection{Strain}
Strain can alter the spin transition frequencies of donors in silicon through several mechanisms. It has been shown that the nuclear magnetic resonance (NMR) frequencies of donors with nuclear spin $I > 1/2$ \cite{FraPRL15, FraPRB16} (e.g. arsenic, antimony and bismuth) can be shifted through a strain-induced quadrupole interaction (QI). Strain can also modify the hyperfine interaction strength $A$ \cite{DrePRL11} or the electron $g$-factor $g_{\rm e}$, both resulting in shifts of the spin resonance frequencies. Here we will analyze all three of these mechanisms (QI, $A$ and $g_{\rm e}$) to determine if they are capable of accounting for the ESR spectra presented in Section~\ref{Sec:CFS}.

In order to aid in our discussion, we first explain the origin of strain in our device and provide an estimate of its magnitude and spatial distribution through simulations. The aluminium resonator is deposited on the silicon substrate by electron-beam evaporation at room temperature, where the device is assumed to be strain-free \cite{EipPD04, MadCRC02}. Whilst the evaporation temperature may be above room temperature in practice, it is assumed to be only a fraction of the total temperature range explored in our experiments ($\Delta T \approx 300$~K). As the device is cooled to 20~mK, the approximate ten-fold difference in the CTEs of silicon and aluminium produces substantial device strains through differential thermal contraction. We perform finite-element simulations of these strains using the software package COMSOL Multiphysics, where we include temperature-dependent CTEs of the materials \cite{SweJPCRD83,NixPR41,FanSAA00} and the anisotropic stiffness coefficients for silicon \cite{HopIEEE10}. Three of the six independent strain tensor components (those along the $\left<100\right>$ crystal axes) have been plotted in Fig.~\ref{fig:strain} as a function of position. The full strain tensor and its spatial dependence can be found in Appendix~\ref{app:STS}.

\begin{figure}
\includegraphics[width=86mm]{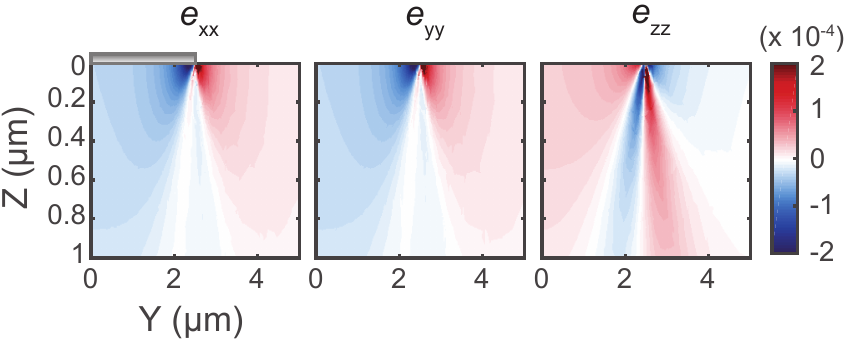}
\caption{\label{fig:strain} Finite-element COMSOL simulations of the strain tensor components along the principle crystal axes x~$\parallel\left[100\right]$, y~$\parallel\left[010\right]$ and z~$\parallel\left[001\right]$ and their variation as a function of position in the device. A cross-section of the aluminium wire (drawn to scale) is represented by the grey gradient-filled box above the silicon substrate (which bounds the strain data). Only half of the wire is displayed here due to it being symmetric about its center. Here the xyz crystal axes are related to the sample frame XYZ (used to describe $\delta B_{\rm 1}$ and the orientation of $B_{\rm 0}$) by a $45^\circ$ rotation about Z (see Fig.~\ref{fig:device}a). We show the result for the wire running parallel to the $\left[110\right]$ (or X) axis, the direction in which the sample was cleaved.}
\end{figure}

\subsubsection{Quadrupole Interaction}
There have been several recent studies that report on quadrupole interactions of group-V donors in silicon, generated by strain \cite{FraPRL15, FraPRB16} or interface defects \cite{MorNT16}. Nuclei with a spin $I > 1/2$ can have a non-spherical charge distribution and associated with this is a quadrupole moment $\mathcal{Q}$ \cite{KauRMP79}. This charge distribution has an axis of symmetry that aligns with the nuclear angular momentum and interacts with an electric field gradient (EFG) $V_{\rm\alpha\beta}$ (where $\alpha$ and $\beta$ are principal axes in the local crystal coordinate system) produced by external charges, such as the donor-bound electron. The interaction is described by the following quadrupole Hamiltonian:

\begin{equation}
\label{eq:QuadHam}
H_{\rm Q}/h = \gamma\frac{e\mathcal{Q}V_{\rm zz}}{4I\left(2I-1\right)h}\left[3I^2_{\rm z}-\mathbf{I}^2+\eta\left(I^2_{\rm x}-I^2_{\rm y}\right) \right] 
\end{equation}\\
where $\gamma$ is a multiplicative scaling factor (resulting from the Sternheimer anti-shielding effect \cite{KauRMP79}), $e$ is the electron charge, $h$ is Planck's constant, $\mathbf{I}$ is the nuclear spin operator with components $I_{\rm \alpha}$, $I$ in the denominator is the scalar value of the nuclear spin ($I = 9/2$) and $\eta = \left(V_{\rm xx} - V_{\rm yy}\right)/V_{\rm zz}$ is an asymmetry parameter. It is evident from Eq.~\ref{eq:QuadHam} that the existence of an EFG $V_{\rm\alpha\beta}$ produces a frequency shift between states with different nuclear spin projections $m_{\rm I}$. In the case of the Si:Bi spin system, quadrupole shifts in the ESR spectra are evident at low magnetic fields because the electron and nuclear spin states are strongly mixed by the hyperfine interaction.


In Table~\ref{table:trans} we list the sensitivities of the transitions to the quadrupole coefficient $Q_{\rm zz} = \gamma e\mathcal{Q}V_{\rm zz}/\left[4I(2I-1)h\right]$ (the prefactor in the quadrupole Hamiltonian $H_{\rm Q}$). By comparing the sensitivities $df/dQ$ to the extended ESR spectra (Fig.~\ref{fig:extdspec}) and observed peak splittings (Table~\ref{table:extdspec}), it becomes apparent that the quadrupole interaction is unlikely to be the origin of the non-trivial spectra shape. The peak splittings $\Delta B_{\rm 0}$ of different transitions for the same resonator approximately follows their magnetic field sensitivities $df/dB_{\rm 0}$ (see Table~\ref{table:trans}), implying an underlying mechanism with a constant frequency distribution across all transitions. This is clearly not the case for the quadrupole interaction, where $df/dQ$ increases with transition number. Furthermore, the $\Delta F\Delta m_{\rm F} = 0$ transitions have sensitivities of opposite sign to the $\Delta F\Delta m_{\rm F} = -1$ transitions -- the asymmetry of this resonance is therefore expected to be opposite that of the low-field peak in the $\Delta F\Delta m_{\rm F} = -1$ transition, as they both correspond to spins in the same region of the device (underneath the wire). However, this is not apparent in Fig.~\ref{fig:extdspec}.

Whereas $df/dQ$ is strongly dependent on the transition, we note that $df/dA$ is constant (see Table~\ref{table:trans}) so that a strain-induced inhomogeneous hyperfine interaction is likely to have the desired properties for the comparison of different transitions.

\subsubsection{Hyperfine Interaction}
\begin{figure}
\includegraphics[width=86mm]{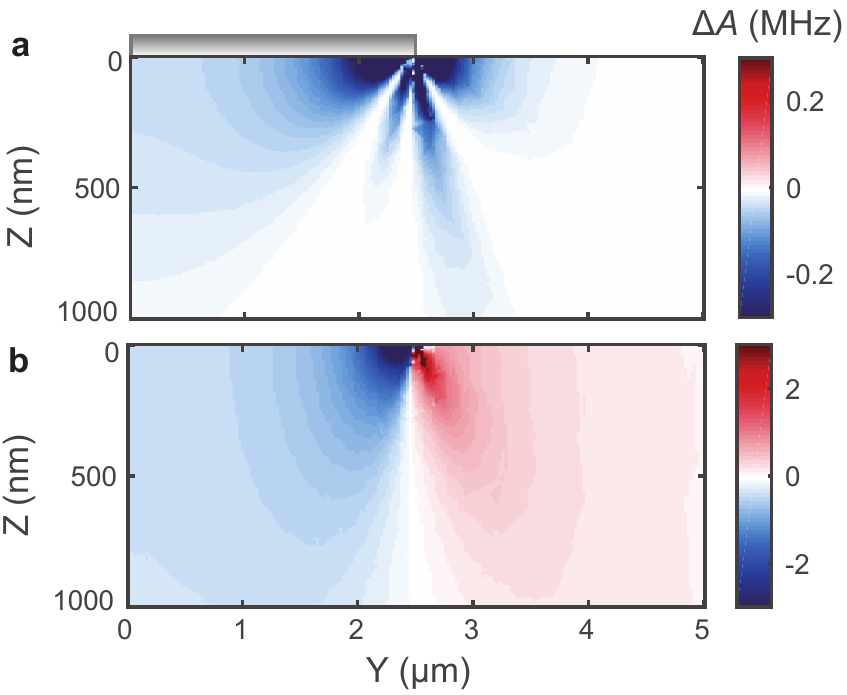}
\caption{\label{fig:hfshift} Calculation of the hyperfine interaction reduction as a result of the simulated device strain. \textbf{(a)} Calculated hyperfine shift $\Delta A$ according to the valley repopulation model, which predicts a quadratic dependence on strain. \textbf{(b)} Calculation performed using the second-order strain model of Eq.~\ref{eq:TB}. The second-order model predicts shifts an order of magnitude larger than the VRM does, as well as displays bipolar frequencies due to its strong linear dependence.}
\end{figure}

Silicon has a conduction band minimum that is six-fold degenerate along the $\left<100\right>$ equivalent crystallographic directions --- commonly referred to as ``valleys'' \cite{AndoRMP82}. The degeneracy of these valleys is broken by the confining potential of the donor, resulting in a singlet A$_{\rm 1}$ ground state and doublet E and triplet T$_{\rm 2}$ excited states \cite{KohnPR55}. For a donor in a bulk silicon crystal (in the absence of strain and electric fields) the electron is perfectly described by the singlet ground state  $\vert\psi\rangle = \vert A_{\rm 1}\rangle$. The E and T$_{\rm 2}$ state wavefunctions have vanishing probabilities at the nucleus (i.e. $\vert\psi(0)\vert^2 = 0$) and consequently do not exhibit a hyperfine interaction ($A = 0$). Applying strain to a valley shifts its energy relative to the conduction band minimum, resulting in a rearrangement of the relative populations of each valley which can be described as a mixing of the donor A$_{\rm 1}$ and E states. The degree of mixing can be calculated using the ``valley repopulation'' model (VRM) \cite{WilPR61}, which predicts a quadratic shift of the hyperfine interaction with an applied strain \cite{ManarXiv17}:
\begin{equation}
\label{eq:VRM}
\begin{aligned}
\frac{\Delta A(\mathbf{\epsilon})}{A(0)} = -\frac{\Xi_{\rm u}^2}{9E_{\rm 12}^2}[(\epsilon_{\rm xx}-\epsilon_{\rm yy})^2+(\epsilon_{\rm xx}-\epsilon_{\rm zz})^2 \\
+(\epsilon_{\rm yy}-\epsilon_{\rm zz})^2]
\end{aligned}
\end{equation}\\
with $\Xi_{\rm u} \approx 8.7$~eV the uniaxial deformation potential of silicon, $E_{\rm 12}$ the energy splitting between the A$_{\rm 1}$ and E states and $\mathbf{\epsilon}$ is a general strain tensor with principal components $\epsilon_{\rm \alpha\alpha}$ (where $\alpha$ are the cubic axes x~$\parallel [100]$, y~$\parallel [010]$ and z~$\parallel [001]$). This expression is valid in the limit of small $\mathbf{\epsilon}$ ($\vert\epsilon_{\rm \alpha\alpha}\vert \lesssim 1\times10^{\rm -3}$) and is applicable for the range of strain produced in our device. In Fig.\ref{fig:hfshift}a we plot the hyperfine shift $\Delta A(\mathbf{\epsilon})$ close to the inductive wire, calculated using Eq.~\ref{eq:VRM}. The quadratic dependence of $A(\mathbf{\epsilon})$ on strain implies that it is only reduced from $A(0)$, the un-strained value. It is apparent that such a distribution could not explain the spectra of Section~\ref{Sec:CFS}, which would require both positive and negative frequency components in order to split the resonance peak in the manner observed. In addition, the VRM predicts $\Delta A \approx 100$~kHz for strains of order $10^{\rm -4}$, equating to a resonance shift of $\Delta A \times (df/dA)/(df/dB_{\rm 0}) = 20~\mu$T, an order of magnitude smaller than our observed peak broadening.

Very recently, it was found that the hyperfine interaction of donors in silicon is also sensitive to the hydrostatic component of strain \cite{ManarXiv17}. This result is surprising, as the VRM predicts no hyperfine reduction for strains that shift all of the valleys by the same energy. A second-order strain model for the hyperfine shift was proposed:
 \begin{equation}
\label{eq:TB}
\begin{aligned}
\frac{\Delta A(\mathbf{\epsilon})}{A(0)} = \frac{K}{3}(\epsilon_{\rm xx}+\epsilon_{\rm yy}+\epsilon_{\rm zz}) + \frac{L}{2}[(\epsilon_{\rm xx}-\epsilon_{\rm yy})^2 \\
+(\epsilon_{\rm xx}-\epsilon_{\rm zz})^2+(\epsilon_{\rm yy}-\epsilon_{\rm zz})^2] + N(\epsilon_{\rm xy}^2+\epsilon_{\rm xz}^2+\epsilon_{\rm yz}^2)
\end{aligned}
\end{equation}\\
with $K = 29$, $L = -9064$ and $N = -225$ the model coefficients for $^{209}$Bi calculated using tight-binding theory and $K = 17.5$ extracted from a first principles calculation using density functional theory (DFT). Remarkably, for $\vert\epsilon\vert\lesssim 1\times10^{\rm -3}$ the model predicts that the linear hydrostatic strain dominates the hyperfine shift. It is suggested that this term is due primarily to strain effects on the central-cell potential, inducing a coupling between the 1s A$_{\rm 1}$ state and higher donor orbital states with the same symmetry. Experiments confirmed the existence of the linear term and the extracted coefficient $K = 19.1$ was in good agreement with theory. A calculation of the hyperfine shift distribution in the device using the full second-order strain model (Eq.~\ref{eq:TB}) is shown in Fig.~\ref{fig:hfshift}b. For strains of order $10^{\rm -4}$, we expect $\Delta A \approx 1$~MHz and an equivalent resonance shift of $\sim 200~\mu$T. In addition, the sensitivity of the resonance frequency to the hyperfine interaction is constant across all spin transitions (see Table~\ref{table:trans}), in agreement with the peak splittings extracted in Table~\ref{table:extdspec}. This mechanism provides bipolar resonance shifts of the correct magnitude and thus constitutes a likely explanation for the spectra of Fig.~\ref{fig:compsweep}c. It should be noted that such a mechanism is not unique to bismuth, a linear hyperfine tuning with strain was observed for all of the group-V donors in silicon \cite{ManarXiv17}.



\subsubsection{g-Factor}
The final mechanism we consider is a strain-induced shift of the electron g-factor $g_{\rm e}$. Strain modifies $g_{\rm e}$ directly (by admixing higher-lying energy bands) and through the valley repopulation effect \cite{WilPR61}. This alters the gyromagnetic ratio $\gamma_{\rm e} = g_{\rm e}\mu_{\rm B}/h$ (where $\mu_{\rm B}$ is the Bohr magneton), shifting the spin resonance frequency through the electron Zeeman interaction $\gamma_{\rm e}\mathbf{B_{\rm 0}}\cdot\mathbf{S}$. The g-factor shift for donors in silicon has been predicted and measured to be several orders of magnitude smaller than that of the hyperfine interaction \cite{WilPR61, BraPRL06}. In addition, the electron Zeeman energy for the range of fields applied in our study ($B_{\rm 0} < 7$~mT) is small, with $E_{\rm z}/h = \gamma_{\rm e}B_{\rm 0} < 300$~MHz, thus providing a proportionally lower contribution to the transition frequency than the hyperfine interaction $A = 1457$~MHz. We quantify this with the transition parameter data in Table~\ref{table:trans}. For the same relative change, the hyperfine interaction shifts the resonant frequency by a factor $(A\times df/dA)/(g_{\rm e}\times df/dg)$  greater than does the electron $g$-factor, which ranges from 50-100 for the spin transitions explored here. Finally, comparing the g-factor sensitivity $df/dg$ for the transitions of resonator A to those of resonator B, we expect the splittings and broadenings to be a factor $\sim$~2 larger for resonator B, which is not observed in the measurements. We therefore safely neglect this mechanism.

\section{ESR Spectra Simulations}\label{Sec:SS}
\begin{figure}
\includegraphics[width=86mm]{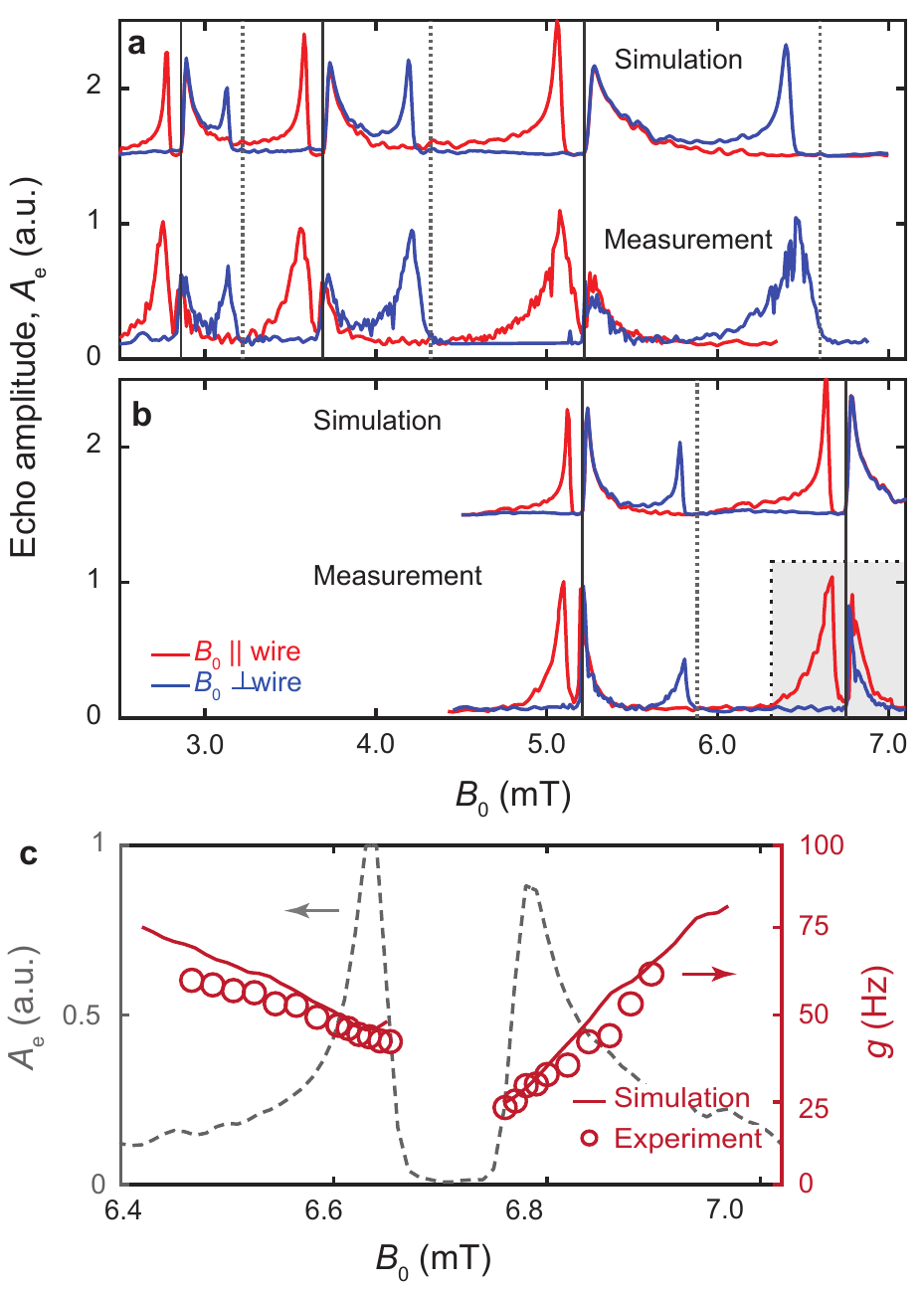}
\caption{\label{fig:spectrasim} \textbf{(a)} Compensated echo-detected field sweeps of transitions 1A-8A (resonator A). The bottom traces are measured data from Fig.~\ref{fig:extdspec} (plotted again here for ease of comparison with theory), whilst the top traces have been offset intentionally and are the results of our theoretical modeling. \textbf{(b)} Compensated echo-detected field sweeps (measurement and simulation) of transitions 1B-4B (resonator B). \textbf{(c)} The single spin-resonator coupling strength $g$ as a function of field $B_{\rm 0}$, extracted from transition 4B (marked by the black dashed box in panel b). The red open circles are derived from measurements of the Rabi frequency. Quantitative agreement is observed with the simulated data (red solid line).}
\end{figure}

In this section we assess whether the hydrostatic hyperfine shift can reproduce the measurement data by performing a full simulation of the extended ESR spectra of Fig.~\ref{fig:extdspec}. The upper offset traces of Figs.~\ref{fig:spectrasim}a and \ref{fig:spectrasim}b are the results of a numerical model incorporating the finite-element simulation of $\delta B_{\rm 1}$ and the hyperfine shift calculations (found by applying Eq.~\ref{eq:TB} to the strain simulations of Fig.~\ref{fig:strain}). For every pixel in the device where dopants are present, we use the pre-determined $\Delta A(\mathbf{\epsilon})$ (with the experimental value of $K$) and calculate the spin transition parameters by solving the modified Hamiltonian:

\begin{align}
H/h &= H_{\rm 0}/h + \Delta A(\epsilon)\mathbf{S}\cdot\mathbf{I} \label{eq:FullHamA}
\end{align}\\

At each $B_{\rm 0}$ we calculate the spectral overlap of all allowed transitions ($\Delta m_{\rm F} = \pm 1$ and $\Delta m_{\rm F} = 0$) with the resonator and weight the resulting spectrum from each pixel with the corresponding donor concentration and the appropriate component of the magnetic field vacuum fluctuations (Fig.~\ref{fig:B1sim}), summing over all pixels to achieve the spectra in Figs.~\ref{fig:spectrasim}a and b. We note that the donor doping profile used in this model (red solid curve in Fig.~\ref{fig:device}b) is the output of a TCAD simulation (discussed in Section~\ref{Sec:PSM}) that takes into account the ionization of donors in the depletion region of the Schottky junction formed between the aluminium wire and the silicon substrate. The simulation strikingly reproduces many features in the experimental data, including peak splittings, peak-height asymmetries and field orientation $\phi$ dependence, without a single free parameter in the model.

Having successfully reproduced key features of the ESR spectra, we investigate whether our model can also capture the correlation of the magnetic field vacuum fluctuations $\delta B_{\rm 1}$ and spin resonance frequency, as discussed in Section~\ref{Sec:CFS}. As noted previously, the Rabi frequency can be expressed in terms of the single spin-resonator coupling strength $g = \gamma_{\rm e}M\vert\delta B_{\rm 1\perp}\vert$ and the mean intra-cavity photon number $\overline{n}$ through the relation $\Omega_{\rm R} = 2g\sqrt{\overline{n}}$. In the compensated sweeps $\Omega_{\rm R}$ is held constant as we pass over the transitions. We extract $g$ as a function of $B_{\rm 0}$ for transition 1B (identified by a black dashed box in Fig.~\ref{fig:spectrasim}b) by estimating $\overline{n}$ at each field using the experimental input power and a calibration of the loss in our setup \cite{BienN16}. In Fig.~\ref{fig:spectrasim}c we plot the result of the experiment (red open circles) overlaid on the simulated spectra (grey dashed line). The data quantifies the qualitative description offered earlier: the coupling strength (or equivalently the vacuum fluctuations $\delta B_{\rm 1}$) increases for the spins that are further detuned (those close to the edge of the wire) and reduces towards the center of the transition, reaching the lowest couplings at the inner-edge of the high-field peak (the spins farthest from the wire). Next, we use our model to simulate the expected $g$ versus $B_{\rm 0}$ dependence, the result (red solid line in Fig.~\ref{fig:spectrasim}c) is an almost quantitative match to the experimental data. 

\section{Summary}\label{Sec:S}
We discussed a range of mechanisms capable of altering the resonance frequencies of donors in micro and nanoelectronic devices and found that strain resulting from differential thermal contraction plays a considerable part. We presented a technique to study such strains in silicon devices through high-sensitivity orientation-dependent ESR spectroscopy. Our results are quantitatively reproduced by considering the shift of the hyperfine interaction caused by the hydrostatic component of strain \cite{ManarXiv17}. The resulting resonance frequency shifts of $\sim~5$~MHz for strains of $\sim 10^{-4}$ contributed to an order-of-magnitude broadening of the ESR lines. Whilst the measurements were performed on bismuth donors in silicon, similar effects are expected for the other group-V donors \cite{ManarXiv17}.

The level of agreement demonstrated between our model, which combined finite-element simulations and experimentally-determined Hamiltonian parameters, with the measured data shows that it accurately captures the underlying physics. Remarkably, the simulation quantitatively reproduces the experimental results with no free parameters in the model. This analysis could be adapted to other device geometries and spin systems, and may prove to be useful for spin-based device design. The results presented in this work have implications for QIP with donors and in hybrid systems such as superconducting quantum memories, which require predictability of spin resonance frequencies and the ability to engineer narrow spin line-widths.

The high sensitivity of the donor hyperfine interaction to hydrostatic strain could be used to create a sensitive local probe for strain in nanoelectronic devices. We estimate that with typical intrinsic line-widths achieved for donors in isotopically enriched silicon of $\sim 2$~kHz \cite{MuhNN14}, a single donor could be used to measure strains below $10^{-7}$. This could be integrated with other techniques for donor metrology \cite{MohNL13} to provide valuable insight into the spatial variation of physical system parameters in nanoscale quantum devices. The large strain sensitivity also opens the prospect of driving spin resonance via mechanical resonators, or coupling donors to phonons in circuit quantum electrodynamics experiments.

\begin{acknowledgments}
\emph{Acknowledgments} We thank B. Lovett and P. Mortemousque for fruitful discussions. We acknowledge the support of the European Research Council under the European Community's Seventh Framework Programme (FP7/2007-2013) through grant agreements No. 615767 (CIRQUSS), 279781 (ASCENT) and 630070 (quRAM), and of the Agence Nationale de la Recherche (ANR) through the project QIPSE. J.J.L.M. was supported by the Royal Society. T.S. was supported by the US Department of Energy under contract DE-AC02-05CH11231. F.A.M. and A.M. were supported by the Australian Research Council Discovery Project DP150101863. We acknowledge support from the Australian National Fabrication Facility.
\end{acknowledgments}

\appendix
\section{Spin Resonance Transitions} 
\label{app:SRT}
The hyperfine interaction $A\mathbf{S}\cdot\mathbf{I}$ couples states in the $\vert m_{\rm S}, m_{\rm I}\rangle$ basis that differ in the electron and nuclear spin projections such that $\Delta m_{\rm S} = \pm 1$ and  $\Delta m_{\rm I} = \mp 1$. This can be seen by rewriting the interaction as a product of the spin raising and lowering operators:
\begin{align}
A\mathbf{S}\cdot\mathbf{I} =& A\left(S_{\rm X}I_{\rm X}+S_{\rm Y}I_{\rm Y}+S_{\rm Z}I_{\rm Z}\right)\\ 
=& A\left(S_{\rm Z}I_{\rm Z}+\frac{1}{2}[S_{\rm +}I_{\rm -}+S_{\rm -}I_{\rm +}]\right)\notag
\end{align}
In the coupled $\vert F, m_{\rm F}\rangle$ basis, these states therefore share the same value of $m_{\rm F} = m_{\rm S} + m_{\rm I}$. In general, we can expand the $\vert F, m_{\rm F}\rangle$ basis on the $\vert m_{\rm S}, m_{\rm I}\rangle$ basis as:
\begin{equation}\label{eq:FmF}
\vert F_{\rm \pm}, m_{\rm F}\rangle = a^\pm_{\rm m_{\rm F}}\vert\pm\frac{1}{2}, m_{\rm F}\mp\frac{1}{2}\rangle + b^\pm_{\rm m_{\rm F}}\vert\mp\frac{1}{2}, m_{\rm F}\pm\frac{1}{2}\rangle
\end{equation}
where we use $F_{\rm \pm}$ to represent the higher or lower multiplet $F_{\rm \pm} = I \pm S$ (i.e. $F_{\rm +} = 5$ and $F_{\rm -} = 4$ for $^{209}$Bi or the triplet and singlet states for $^{31}$P). This is true for all states aside from those with $m_{\rm F} = \pm(I+S)$ (corresponding to $\vert m_{\rm S} = \pm S, m_{\rm I} = \pm I\rangle$), which remain unmixed. The mixing coefficients $a^\pm_{\rm m_{\rm F}}$ ($a^+_{\rm m_{\rm F}} = a^-_{\rm m_{\rm F}}$) and $b^\pm_{\rm m_{\rm F}}$ ($b^+_{\rm m_{\rm F}} = -b^-_{\rm m_{\rm F}}$) are determined by the value of $m_{\rm F}$, the hyperfine interaction strength $A$ and the external magnetic field $B_{\rm 0}$ (or more precisely, the electron Zeeman energy relative to the hyperfine interaction) \cite{MohPRB10}. At high magnetic fields (where $E_{\rm z}/h = \gamma_{\rm e}B_{\rm 0}\gg A$) $a^\pm_{\rm m_{\rm F}} \rightarrow 1$ and $b^\pm_{\rm m_{\rm F}} \rightarrow 0$, whilst at low magnetic fields (where $\gamma_{\rm e}B_{\rm 0}\lesssim A$) strong mixing occurs.

\subsection{``$S_{\rm X}$'' Type}
When operating in the ``orthogonal mode'' of spin resonance ($B_{\rm 1} \perp B_{\rm 0}$), the $B_{\rm 1}$ field couples to the $S_{\rm X}$ and $I_{\rm X}$ spin operators. Electron spin resonance transitions may be driven between $\vert F, m_{\rm F}\rangle$ states that contain components of the uncoupled basis that differ by $\Delta m_{\rm S} = \pm 1$, i.e. $\vert F_{\rm \pm}, m_{\rm F}\rangle \leftrightarrow \vert F_{\rm \pm}, m_{\rm F}-1\rangle$ and $\vert F_{\rm \pm}, m_{\rm F}\rangle \leftrightarrow \vert F_{\rm \mp}, m_{\rm F}-1\rangle$, as can be seen from Eq.~\ref{eq:FmF}. The first two transitions ($\vert F_{\rm +}, m_{\rm F}\rangle \leftrightarrow \vert F_{\rm +}, m_{\rm F}-1\rangle$ and $\vert F_{\rm -}, m_{\rm F}\rangle \leftrightarrow \vert F_{\rm -}, m_{\rm F}-1\rangle$) correspond to high-field NMR transitions (which become ESR-allowed at low fields), whilst the third transition ($\vert F_{\rm +}, m_{\rm F}\rangle \leftrightarrow \vert F_{\rm -}, m_{\rm F}-1\rangle$) is a high-field ESR transition and the fourth ($\vert F_{\rm -}, m_{\rm F}\rangle \leftrightarrow \vert F_{\rm +}, m_{\rm F}-1\rangle$) is completely forbidden at high fields -- it corresponds to transitions where $\Delta m_{\rm S} = \pm 1$ and  $\Delta m_{\rm I} = \mp 2$.

The transition matrix elements between these states are given by:
\begin{subequations}\label{eq:MEX}
\begin{widetext}
\begin{align}
\langle F_{\rm \pm}, m_{\rm F} \vert S_{\rm X} + \delta I_{\rm X} \vert F_{\rm \pm}, m_{\rm F}-1\rangle = &\left[ a^+_{\rm m_{\rm F}}b^+_{\rm m_{\rm F}-1}+\delta(a^+_{\rm m_{\rm F}}a^+_{\rm m_{\rm F}-1}+b^+_{\rm m_{\rm F}}b^+_{\rm m_{\rm F}-1})\right]/2,\\
&\left[b^-_{\rm m_{\rm F}}a^-_{\rm m_{\rm F}-1}+\delta(a^-_{\rm m_{\rm F}}a^-_{\rm m_{\rm F}-1}+b^-_{\rm m_{\rm F}}b^-_{\rm m_{\rm F}-1})\right]/2\\
\langle F_{\rm \pm}, m_{\rm F}\vert S_{\rm X} + \delta I_{\rm X} \vert F_{\rm \mp}, m_{\rm F}-1\rangle = &\left[a^+_{\rm m_{\rm F}}a^-_{\rm m_{\rm F}-1}+\delta(a^+_{\rm m_{\rm F}}b^-_{\rm m_{\rm F}-1}+b^+_{\rm m_{\rm F}}a^-_{\rm m_{\rm F}-1})\right]/2,\\
&\left[b^-_{\rm m_{\rm F}}b^+_{\rm m_{\rm F}-1}+\delta(b^-_{\rm m_{\rm F}}a^+_{\rm m_{\rm F}-1}+a^-_{\rm m_{\rm F}}b^+_{\rm m_{\rm F}-1})\right]/2
\end{align}
\end{widetext}
\end{subequations}
where $\delta = \gamma_n/\gamma_e$ is the ratio of the nuclear and electron spin gyromagnetic ratios, which is typically of order $10^{-4}$ for group-V donors in silicon. At low and intermediate fields ($\gamma_{\rm e}B_{\rm 0}\lesssim A$), the first term in the matrix elements of Eqs.~\ref{eq:MEX}a-d  dominate over the components generated by the nuclear spin (those multiplied by $\delta$). At high magnetic fields, the nuclear spin component of the matrix element is negligible for Eq.~\ref{eq:MEX}c but is the dominant term in Eqs.~\ref{eq:MEX}a,b (the high-field NMR transitions). It should be noted that, in general, the matrix elements above are non-zero at low-fields, with the exception of identical particles ($S = I$ and $\delta = 1$) where the singlet state ($F = 0$) is ESR inactive. The singlet state becomes ESR active (for example, in the case of phosphorus $S = I = 1/2$) due to the differing gyromagnetic ratios of the electron and nuclear spins.

\subsection{``$S_{\rm Z}$'' Type}
In the ``parallel mode'' of spin resonance ($B_{\rm 1} \parallel B_{\rm 0}$), the $B_{\rm 1}$ field couples to the $S_{\rm Z}$ and $I_{\rm Z}$ spin operators. Electron spin resonance transitions may be driven between $\vert F, m_{\rm F}\rangle$ states that contain identical components of the uncoupled basis, i.e. $\vert F_{\rm \pm}, m_{\rm F}\rangle \leftrightarrow \vert F_{\rm \mp}, m_{\rm F}\rangle$ (see Eq.~\ref{eq:FmF}). These correspond to high-field flip-flop transitions ($\Delta m_{\rm S} = \pm 1$ and  $\Delta m_{\rm I} = \mp 1$). The matrix element between these states is given by:
\begin{align}\label{eq:MEZ}
&\langle F_{\rm +}, m_{\rm F} \vert S_{\rm Z} + \delta I_{\rm Z} \vert F_{\rm -}, m_{\rm F}\rangle = (a^+_{\rm m_{\rm F}}b^-_{\rm m_{\rm F}}-a^-_{\rm m_{\rm F}}b^+_{\rm m_{\rm F}})/2\notag\\
&+\delta[a^+_{\rm m_{\rm F}}b^-_{\rm m_{\rm F}}(m_{\rm F}-1/2)+a^-_{\rm m_{\rm F}}b^+_{\rm m_{\rm F}}(m_{\rm F}+1/2)]\notag\\
&=-a^+_{\rm m_{\rm F}}b^+_{\rm m_{\rm F}}+\delta a^+_{\rm m_{\rm F}}b^+_{\rm m_{\rm F}}
\end{align}
where we have used the symmetry of the mixing coefficients ($a^+_{\rm m_{\rm F}} = a^-_{\rm m_{\rm F}}$ and $b^+_{\rm m_{\rm F}} = -b^-_{\rm m_{\rm F}}$) to arrive at the final form of Eq.~\ref{eq:MEZ}. Note, for identical gyromagnetic ratios ($\delta = 1$) the components of the matrix element cancel exactly, and driving in the parallel mode is forbidden. Furthermore, at high fields (where $a^\pm_{\rm m_{\rm F}} \rightarrow 1$ and $b^\pm_{\rm m_{\rm F}} \rightarrow 0$) the matrix element becomes negligibly small. For the Si:Bi system ($\delta \approx 10^{-4}$) at low magnetic fields, the ``$S_{\rm Z}$'' transitions are appreciable, comparable in strength to the ``$S_{\rm X}$'' type.

\section{Predicted ESR Line Shape} 
\label{app:LBM}
Previous studies of the sample utilized in this work \cite{WeiAPL12}, performed using a bulk ESR spectrometer (i.e. without the on-chip resonator) revealed a Gaussian line-shape with a peak-to-peak width of $\sigma_{\rm B} = 12~\mu$T for the high-field $m_{\rm I} = -1/2$ transition ($\vert 4, -1\rangle \leftrightarrow \vert 5, 0\rangle$ in the $\vert F, m_{\rm F}\rangle$ basis), at a frequency of $\omega/2\pi = 9.53$~GHz. This transition displays a $df/dB = 0.6\gamma_{\rm e}$ and thus an equivalent $\sigma_{\rm f} = \sigma_{\rm B}\times df/dB_{\rm 0} = 200$~kHz broadening in the frequency domain. This value agrees well with other studies of bismuth-doped isotopically enriched silicon \cite{WolfNN13}, where a line-width of 270~kHz was measured and found to be constant in the frequency domain. For the $\vert 4, -4\rangle \leftrightarrow \vert 5, -5\rangle$ transition studied in this work (with $\omega/2\pi \approx 7.3$~GHz), $df/dB_{\rm 0} = 0.9\gamma_{\rm e}$ (see Table~\ref{table:trans}) and we expect a $\sigma_{\rm B} = \sigma_{\rm f}/(df/dB_{\rm 0}) = 8~\mu$T, providing a full-width-at-half-maximum (FWHM) of $\sim~20~\mu$T. This is substantially lower than the broadening we observe in our measurements using the on-chip micro-resonator, as depicted in Fig.~\ref{fig:compsweep}.

\section{Magnetic Field Inhomogeneity} 
\label{app:MFI}
\begin{figure}[h]
\includegraphics[width=86mm]{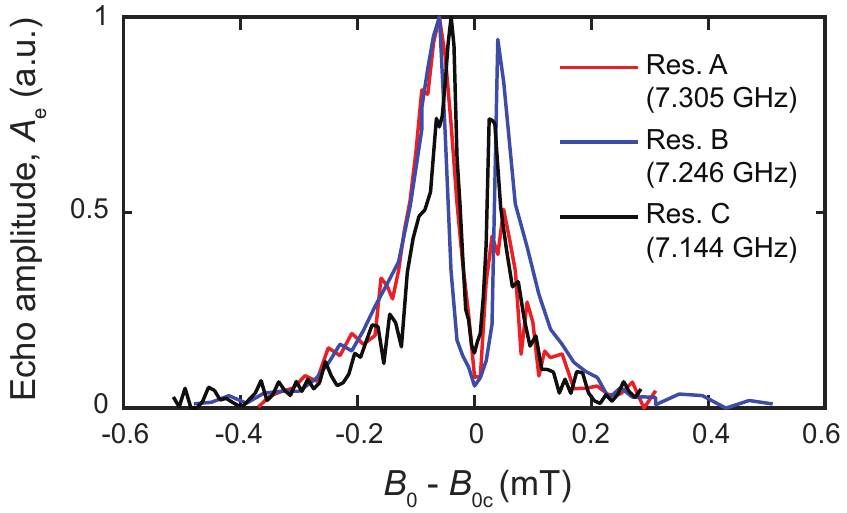}
\caption{\label{fig:firsttranscomp} Compensated echo-detected magnetic field sweep (see Section~\ref{Sec:CFS} for measurement details) recorded for three resonators (frequencies are listed in the figure legend) over the first spin transition $\vert 4, -4\rangle \leftrightarrow \vert 5, -5\rangle$. The horizontal axis displays the difference with the transition center fields $B_{\rm 0c}$ (listed in Table~\ref{table:extdspec}). The transition frequency has a magnetic field sensitivity of $df/dB_{\rm 0} = -0.9\gamma_{\rm e}$ for all three resonators.}
\end{figure}

A magnetic field inhomogeneity, for example produced by Meissner screening of the static magnetic field $B_{\rm 0}$ in the vicinity of the superconducting wire, is not suspected to contribute to the splitting and broadening of the electron spin resonance (ESR) peaks observed in our experiment (Figs.~\ref{fig:compsweep} and \ref{fig:extdspec}). We rule this mechanism out by comparing measurements of the first spin resonance transition $\vert 4, -4\rangle \leftrightarrow \vert 5, -5\rangle$ (see Table~\ref{table:trans}) for each of the three resonators A, B and C (see Fig.~\ref{fig:firsttranscomp}). The width and splitting of these peaks are of similar size for each resonator, despite the transition for resonator C ($B_{\rm 0} = 9.29$~mT, $df/dB_{\rm 0} = -0.90\gamma_{\rm e}$) occurring at twice the field of resonator B and three times the field of resonator A. A broadening resulting from an inhomogeneous magnetic field would increase in proportion to the strength of the field.

\section{Strain Tensor Simulation}
\label{app:STS}
\begin{figure}[b]
\includegraphics[width=86mm]{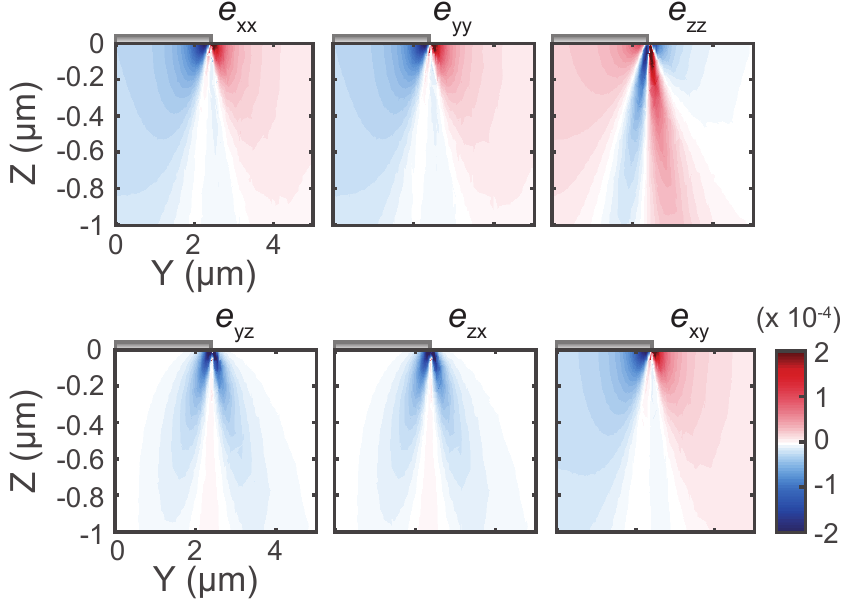}
\caption{\label{fig:fullstraintensor} Spatial dependence of the six independent components of strain in the silicon substrate. Donors are implanted up to a depth of $\sim$~300~nm. The strain tensor components are given in the cubic $\left[100\right]$ basis (xyz), whilst the cut through the device is such that X~$\parallel \left[110\right]$ and Y~$\parallel \left[\overline{1}10\right]$.}
\end{figure}

We have performed finite-element strain simulations of our device using the software package COMSOL Multiphysics. Our model consists of a 50~nm thick, 5~$\mu$m wide aluminium wire on a silicon substrate. We assume the aluminium to be strain-free upon deposition and we simulate cooling the device to 20 mK using the temperature-dependent CTE of aluminium \cite{NixPR41,FanSAA00} and silicon \cite{SweJPCRD83}, as well as the anisotropic stiffness coefficients for silicon \cite{HopIEEE10}. The wire is constructed at a $45^\circ$ degree angle to the x-axis in the xy-plane (where x~$\parallel \left[100\right]$) such that it is aligned with the $\left[110\right]$ crystal axis. The difference in the CTE of Si and Al produces device strains at low temperature. At each pixel in the device we extract the six independent strain components in the $\left<100\right>$ basis, which are plotted in Fig.~\ref{fig:fullstraintensor}.

\bibliography{Strain}
\end{document}